\documentclass[twoside,12pt]{article}
\usepackage{latexsym,amsmath,amssymb,url,amsthm}
\usepackage{wrapfig}
\usepackage{tikz}
\def\,{\mskip 3mu} \def\>{\mskip 4mu plus 2mu minus 4mu} \def\;{\mskip 5mu plus 5mu} \def\!{\mskip-3mu}
\def\dispmuskip{\thinmuskip= 3mu plus 0mu minus 2mu \medmuskip=  4mu plus 2mu minus 2mu \thickmuskip=5mu plus 5mu minus 2mu}
\def\textmuskip{\thinmuskip= 0mu                    \medmuskip=  1mu plus 1mu minus 1mu \thickmuskip=2mu plus 3mu minus 1mu}
\textmuskip
\def\beqn{\dispmuskip\begin{displaymath}}\def\eeqn{\end{displaymath}\textmuskip}
\newenvironment{keywords}{\centerline{\bf\small
Keywords}\begin{quote}\small}{\par\end{quote}\vskip 1ex}
\def\paradot#1{\vspace{1.3ex plus 0.5ex minus 0.5ex}\noindent{\bf{#1.}}}

\begin{document}
\title{
\vskip 2mm\bf\Large\hrule height5pt \vskip 4mm
Adaptive Context Tree Weighting
\vskip 4mm \hrule height2pt
}

\author{{\bf Alexander O'Neill}$^1$, {\bf Marcus Hutter}$^{1,2}$, \\ {\bf Wen Shao}$^1$, {\bf Peter Sunehag}$^1$\\[3mm]
\normalsize Research School of Computer Science \\[-0.5ex]
\normalsize $^1$Australian National University and $^2$ETH Z{\"u}rich \\[-0.5ex]
\normalsize\texttt{\{marcus.hutter, wen.shao, peter.sunehag\}@anu.edu.au}
}
\date{January 2012}
\maketitle

\begin{abstract}
We describe an adaptive context tree weighting (ACTW)
algorithm, as an extension to the standard context tree
weighting (CTW) algorithm. Unlike the standard CTW algorithm,
which weights all observations equally regardless of the depth,
ACTW gives increasing weight to more recent observations,
aiming to improve performance in cases where the input sequence
is from a non-stationary distribution.
Data compression results show ACTW variants improving over CTW
on merged files from standard compression benchmark tests while
never being significantly worse on any individual
file.\footnote{The results in this article first appeared in
the honour's thesis of the first author \cite{Nei10}.}
\\
\def\contentsname{\centering\normalsize Contents}
{\parskip=-2.7ex\tableofcontents}
\end{abstract}

\begin{keywords}
Data compression;
universal code;
prediction;
Context Tree Weighting (CTW) algorithm.
\end{keywords}

\newpage
\section{Introduction} \label{sec:introduction}

Data compression is the task of encoding a data source into a
more compact representation. In this paper, we are mainly
interested in the task of lossless data compression, which
requires that the original data must be exactly reproducible
from the compressed encoding. There are a number of different
techniques for lossless data compression. Some of the more
popular methods employed include Burrows-Wheeler transform
encoders \cite{BW94}, those based on Lempel-Ziv coding
\cite{LZ77,LZ78}, using Dynamic Markov compression (DMC)
\cite{CH87} or using prediction by partial matching (PPM)
\cite{CW84}. Many data compressors make use of a concept called
arithmetic coding \cite{Ris76,LR79,Wit87}, which when provided
with a probability distribution for the next symbol can be used
for lossless compression of the data. In general, however, the
true distribution for the next symbol is unknown and must be
estimated. For stationary distributions it is easy to estimate
the true distribution, which makes arithmetic coding
asymptotically optimal. For non-stationary distributions this
is no longer the case, and it is this problem we tackle in this
paper.

CTW is an online binary prediction algorithm first presented in
\cite{WST95}. The general idea is to assume bits are generated
by an unknown prediction suffix tree \cite{RST96} of bounded
depth and then use a Bayesian mixture to perform prediction.
The CTW algorithm does this by means of a weighted context
tree, which efficiently represents the Bayesian mixture over
all suffix trees of bounded depth. This allows CTW to perform
updates in time that grows linearly with depth, whereas a naive
approach would lead to double exponential time.
As such, it is an efficient, general purpose sequence
prediction method that has been shown to perform well both
theoretically and in practice \cite{BEY04}.

It has been shown that CTW can be applied to lossless data
compression \cite{WST97}. The standard CTW algorithm, however,
was designed specifically for coding sequences from a
stationary source, so it is not surprising that it may perform
poorly when the sequence is generated by a non-stationary (e.g.
drifting) source. In this paper, we address this problem by
introducing the adaptive CTW algorithm, which quickly adapts to
the current distribution by increasing the weight of more
recent symbols. It has been shown in \cite{Nei10} that this
adaptive CTW algorithm can improve results when integrated with
a Monte Carlo AIXI agent\cite{VNHUS11}.

\paradot{Structure of this paper}
This paper begins with a brief description of the standard CTW
algorithm (Section 2). We then introduce several variants of
ACTW algorithm (Section 3). Experimental results as well as an
analysis of the performance of the ACTW variants are given in
Section 4. Conclusions are drawn in the final section.

\section{The CTW algorithm} \label{sec:background}

The CTW algorithm \cite{WST95,WST97} is a theoretically
well-motivated and efficient online binary sequence prediction
algorithm. It uses Bayesian model averaging that computes a
mixture over all prediction suffix trees \cite{RST96} up to a
certain depth, with greater prior weight given to simpler
models.

\paradot{Krichevsky-Trofimov estimator}
The KT estimator \cite{KT81} is obtained using a Bayesian
approach by assuming a $(\frac{1}{2},\frac{1}{2})$-Beta prior
on the parameter of a Bernoulli distribution. Let $y_{1:t}$ be
a binary string containing $a$ zeros and $b$ ones. We write
$P_{kt}(a,b)$ to denote $P_{kt}(y_{1:t})$. The KT estimator can
be incrementally calculated by: $P_{kt}(a+1,b) =
\frac{a+1/2}{a+b+1}P_{kt}(a,b)$ and $P_{kt}(a,b+1) =
\frac{b+1/2}{a+b+1}P_{kt}(a,b)$ with $P_{kt}(0,0) = 1$.

\paradot{Context Tree Weighing algorithm}
A context tree of depth $D$ is a perfect binary tree of depth
$D$, with the left edges labelled $1$ and the right edges
labelled $0$. Let $n$ be a node in the context tree and suppose
$y_{1:t}$ is the sequence that has been seen so far and
$[y_{1:t}]_{|n}$ is the sequence of bits in $y_{1:t}$ that end
up in $n$. The counts $a_n$ and $b_n$, corresponding to the
number of zeros and ones in $[y_{1:t}]_{|n}$, are stored at
each node $n$, and are updated as bits in the input sequence
are observed. The KT estimate is calculated at each node $n$
based on the attached counts $a_n$ and $b_n$. Additionally, we
introduce a weighted probability for each node $n$, which can
be recursively calculated by
\beqn
P_w^n(y_{1:t}) =
\begin{cases}
P_{kt}([y_{1:t}]_{|n}) & \quad \text{if $n$ is a leaf node}\\
      \frac{1}{2}P_{kt}([y_{1:t}]_{|n})+\frac{P_w^{n_l}([y_{1:t}]_{|n_l})P_w^{n_r}([y_{1:t}]_{|n_r})}{2} & \quad \text{otherwise}
\end{cases}
\eeqn
where $n_l$ and $n_r$ are left and right children of $n$
respectively. The joint probability for the input sequence is
then given by the weighted probability at the root node.

\section{Adaptive CTW} \label{sec:adaptive_ctw}

\paradot{Limitations of standard CTW}
The CTW algorithm is limited by its use of KT estimators to
approximate the current distribution. This is appropriate if
the true distribution is stationary, but not in the
non-stationary case. The problem is that the KT estimator is
very slow to update once many samples have been collected, so
it cannot quickly learn a change in distribution.

\paradot{KT with a moving window and discounted KT}
Adaptive schemes such as \cite{Wil96} and \cite{SM99} have been
studied. They are more computationally expensive than the
standard KT estimator and we want an adaptive scheme that comes
at no extra cost in computation time or memory.
First we motivate our method by considering the KT estimator
with a moving window. We estimate the probability of the next
bit using a standard KT estimator, but instead of using all the
previous history, we only use the last $k$ bits. Suppose a
sequence is generated by $Bern(\theta)$, with some $0\leq
\theta \leq 1$. There is no nice redundancy bound for all
cases, as one may encounter strings like $0^k1^k0^k\ldots$. We
instead studied the expected redundancy for one bit, which is
\beqn
  R(k;\theta) = \sum_{a=0}^k \left(\begin{array}{c}
    k \\
    a
  \end{array}\right) \theta^a(1-\theta)^{k-a}[\theta\log\frac{k+1}{k-a+\frac{1}{2}} + (1-\theta)\log\frac{k+1}{a+\frac{1}{2}}] - H[\theta],
\eeqn
where $H[\theta]$ is the entropy. This quantity can be upper
bounded by $O(1/k)$, and a lower bound of the same order is
given in \cite[Thm. 1]{Kri98}. A shortcoming of this method is
that one has to keep a history of length $k$. We therefore
consider a discounted version of the KT estimator and apply it
to the standard CTW algorithm.
The discounted KT estimator solves the problem of storing a
history of length $k$ while retaining the desired fixed
effective horizon of the windowing method described above in
the sense that only recent bits significantly affect the
prediction. Let $\gamma \in [0, 1)$ denote the discount rate
and $y_{1:t}$ be a binary string. We store discounted count
$a_t$ and $b_t$, corresponding to the discounted number of
zeros and ones in $y_{1:t}$. After we observe the next symbol,
as with in the standard KT estimator, one of the counts $a_t$
and $b_t$ is incremented according to the observed symbol and
KT estimate is calculated based on $a_t$ and $b_t$. We then use
the discount rate to update the counts $a_t$ and $b_t$ by
\beqn
  a_{t+1} : = (1 - \gamma) \, a_t \qquad b_{t+1} : = (1 - \gamma) \, b_t
\eeqn

\paradot{ACTW}
The adaptive CTW algorithm differs from the standard CTW
algorithm only in the use of the discounted KT estimator.
Therefore the adaptation comes at no extra computation or
memory cost over the standard algorithm.

The value of $ \gamma $ need not be a constant. In the
following we consider a number of possibilities for assigning
$\gamma$.

\paradot{Fixed rate}
The most basic of the adaptive methods implemented is the fixed
rate adaptive method. In this method, the value of $ \gamma $
is fixed, and every update makes use of this same constant.
Consider an observed sequence $ y_{1:t} $, where $
[y_{1:t}]_{\vert n} $ denotes the sequence of bits in $
y_{1:t}$ which end up in $ n $. Let $ k $ be the length of $
[y_{1:t}]_{\vert n} $ and use $ [y_{1:t}]_{\vert n, i} $ to
denote the $ i^{th} $ bit in this sequence. Then for a constant
$ \gamma $ we get a weighting for the $ i^{th} $ bit in $
[y_{1:t}]_{\vert n} $ given by:
\beqn
  w_{n,i} = (1 - \gamma)^{k-i}
\eeqn
Clearly as $ k $ increases the weighting decreases. In the case
of $ \gamma = 0 $ this reduces to the standard CTW algorithm.
For $ \gamma  > 0 $ we have a desired fixed effective horizon
whose length is determined by $\gamma$. While this method is
simple, it suffers from the necessity of choosing parameter
$\gamma$, which determines the fixed effective horizon.

\paradot{Sequence length based}
For this method the discount rate becomes a function of the
length of the sequence observed so far. If $ y_{1:t} $ is the
sequence observed so far, then updates occurring due to the
observation of $ y_{t+1} $ use an discount rate given by
\beqn
  \gamma_{t+1} = c t^{- \alpha} \qquad c, \alpha\in [0,1)
\eeqn
If $ \alpha = 0 $ then this reduces to a fixed-rate adaptive
CTW with $\gamma = c$. For $\alpha > 0$, the adaptive
multiplier decreases over time. Therefore this method leads to
an increasing effective horizon, though at the cost of
decreasing the benefits of adaptivity after observing a long
input sequence.
If there is a long sequence of observations between a node $ n
$ being updated, this method can also lead to significant
variation between weightings assigned to observations at the
node, even if no other observations for that context have been
observed in the meantime.

\paradot{Context visit-based}
To overcome this variation in weightings when dealing with
contexts for which few observations have been made we use a
method where the discount rate becomes a function of how many
times a context has been observed. Below we present three
variations on this idea, which share the same overarching
philosophy but use somewhat different approaches.

\paradot{Partial-context visit-based}
The first approach is the most obvious application of this
principle. At a node $ n $ where $ [y_{1:t}]_{\vert n} $ has
length $ k $ (i.e. $ k $ different bits have been observed that
end up in $ n $), we use an discount rate given by
\beqn
  \gamma_n = c k^{- \alpha} \qquad c, \alpha\in [0,1)
\eeqn
This is perhaps the most elegant approach to the situation
described above. However we note that for nodes higher in the
context tree, which have the greatest impact, the number of
observations of these contexts will increase rapidly, and so
the adaptive multiplier for these nodes will decrease faster.
While this is not necessarily a problem we nevertheless
investigate methods that do not have this property.

\paradot{Full-context visit-based}
In this method we calculate an discount rate at the leaf node
corresponding to the current context as per the previous
method. However, instead of repeating this process at each node
on the path towards the root, we instead propagate the discount
rate up the tree. In this way all nodes on the context path use
the same discount rate calculated at the leaf node. As the
discount rate is based on observations of the leaf node, the
value used for nodes higher in the tree will not be so closely
linked to the sequence length. Formally, when updating a leaf
node $ n $ where $ [y_{1:t}]_{\vert n} $ has length $ k_n $ we
use the discount rate
\beqn
  \gamma_n = c k_n^{- \alpha}
\eeqn
and for each node $ n' $ on the path from $ n $ to the root
node $ \lambda $ we use
\beqn
  \gamma_{n'} = c k_n^{- \alpha}
\eeqn
Therefore the same discount rate is used for all nodes in the
path. Thus for a weighted context tree of depth $ D $ the
discount rate becomes a function of the number of observations
of the current length $ D $ context.

\paradot{Leaf-context visit-based}
This method is similar to the full-context visit-based adaptive
method, but uses an additive rather than multiplicative
approach to updating the counts $ a $ and $ b $. For the leaf
node corresponding to the current context the same discount
rate is used as for the previous two approaches, but for nodes
on the path towards the root we instead update the counts as
the sum of the counts of its child nodes. More formally, when
updating a leaf node $ n $ where $ [y_{1:t}]_{\vert n} $ has
length $ k_n $ we use the discount rate
\beqn
  \gamma_n = c k_n^{- \alpha}
\eeqn
and for nodes $ n' $ on the path from $ n $ to the root node we
update the counts $ a_{n'} $ and $ b_{n'} $ using
\beqn
  a_{n'} = a_{n'_l} + a_{n'_r} \qquad b_{n'} = b_{n'_l} + b_{n'_r}
\eeqn
where $ n'_l $ and $ n'_r $ are the left and right children of
$ n' $ respectively. In this way discount rate has no effect on
the KT-estimator count contributions of any other depth $ D $
context. This approach also preserves the property of CTW where
the counts $ a_n $ and $ b_n $ of a node $ n $ is equal to the
sum of the counts for its child nodes.

\section{Experiments} \label{sec:experiments}

\paradot{Test datasets}
In this section, we evaluate the variants of ACTW against the
standard CTW algorithm\footnote{We used a generic CTW
implementation, not the highly turned one presented in
\cite{TVW97}} across a range of test sets, including the
following standard benchmarks: large calgary corpus
\cite{BWC89}, canterbury
corpus\footnote{\url{http://corpus.canterbury.ac.nz/}} and
single file compression (SFC)
testset\footnote{\url{http://www.maximumcompression.com/index.html}}.
We also tested our algorithm on an assortment of different file
types that were collected for testing compression performance
with changing sources. Details are given in Table
\ref{table:assorted}. The division of files is given in Table
\ref{table:merge-sets}, with the sets concatenated in the order
listed.
\begin{table}[ht!]
  \centering
  \caption{Assorted collection of test files}
  \label{table:assorted}\small
  \begin{tabular}{|r l l|}
    \hline
    \bf Size & \bf Name & \bf Type \\
    \hline
    3639172 & book1.pdf & PDF file \\
    2685309 & book2.txt & ASCII text \\
    656896 & data1.xls & Microsoft Excel spreadsheet \\
    544768 & data2.xls & Microsoft Excel spreadsheet \\
    1841392 & exec1 & UNIX compiled executable \\
    2169915 & exec2.exe & exe executable\\
    6784000 & exec3.msi & msi executable \\
    718377 & flash.swf & Shockwave flash file \\
    345160 & foreign.hwf & foreign language file \\
    561000 & lib.dll & Microsoft Dynamic Link Library \\
    1601949 & pic1.png & PNG image \\
    1861255 & pic2.png & PNG image \\
    55832855 & pitches & pitch values of MIDI files \\
    635392 & pres.ppt & Microsoft PowerPoint presentation \\
    86948 & text.rtf & rich-text format text \\
    2440044 & vid1.avi & avi video file \\
    5167297 & vid2.mov & mov video file \\
    \hline
  \end{tabular}
\end{table}

\begin{table}[ht!]
  \centering
  \caption{Assorted test file merged sets (ordered)}
  \label{table:merge-sets}\small
  \begin{tabular}{|cccc|}
    \hline
    \bf merge1 & \bf merge2 & \bf merge3 & \bf merge4 \\
    \hline
    exec1 & pic1.png & foreign.hwp & data2.xls \\
    vid1.avi & data1.xls & exec3.msi & vid2.mov \\
    flash.swf & exec2.exe & pres.ppt & text.rtf \\
    book1.pdf & book2.txt & pic2.png & lib.dll \\
    \hline
  \end{tabular}
\end{table}

\paradot{Comparison data compressors}
For comparison purposes, we have chosen data compressors to
cover a range of the most commonly used compression techniques,
including LZW, gzip, LZMA, bzip2, PAQ8L.

\paradot{ACTW variants}
We abbreviate different adaptive CTW variants used in experiment as below:
\begin{itemize}\parskip=0ex\parsep=0ex\itemsep=0ex
  \item ACTW1 -  \textit{fixed rate adaptive CTW, $ \gamma = 0.01 $}
  \item ACTW2 -  \textit{partial context visit based adaptive CTW, $ c = 0.1 $, $ \alpha = 0.33 $}
  \item ACTW3 -  \textit{partial context visit based adaptive CTW, $ c = 0.1 $, $ \alpha = 0.5 $}
   \item ACTW4 -  \textit{full context visit based adaptive CTW, $ c = 0.1 $, $ \alpha = 0.33 $}
  \item ACTW5 -  \textit{leaf context visit based adaptive CTW, $ c = 0.1 $, $ \alpha = 0.33 $}
\end{itemize}
The context tree depth for compression testing was set to 28
for all the different CTW and ACTW based compressors. The
parameters are mildly tuned with the objective of not being
significantly worse than CTW on any file. As a consequence we
have more modest gains as well.

\begin{table}[ht!]
  \caption{Compression result, Large Calgary corpus}
  \label{table:calgary-results}
  \resizebox{\textwidth}{!} {
  \begin{tabular}{r|c|c|c|c|c|c|c|c|c|c|c}
  file&LZW&gzip&LZMA&bzip2&PAQ8L&CTW&ACTW1&ACTW2&ACTW3&ACTW4&ACTW5\\
  \hline\hline
  bib&58.18&68.49&72.55&75.31&81.27&69.18&68.81&68.11&68.67&66.75&68.66\\
  book1&58.75&59.24&66.04&69.74&74.92&67.53&67.09&66.94&67.36&66.07&67.29\\
  book2&58.86&66.16&72.21&74.23&80.05&68.55&68.40&68.27&68.53&67.51&68.59\\
  geo&24.05&33.11&48.06&44.41&57.04&35.11&36.03&35.69&35.39&36.01&35.15\\
  news&51.30&61.59&68.49&68.55&76.17&61.83&61.64&61.28&61.62&60.45&61.55\\
  obj1&34.67&51.99&56.33&49.84&65.22&41.85&42.23&42.20&42.12&42.34&41.87\\
  obj2&47.87&66.93&75.13&69.03&81.80&56.65&57.58&57.80&57.34&57.64&57.40\\
  paper1&52.83&65.06&67.58&68.85&75.38&63.16&62.93&62.23&62.68&60.89&62.65\\
  paper2&56.01&63.80&66.92&69.54&75.09&66.11&65.81&65.27&65.75&63.82&65.73\\
  paper3&52.36&61.10&63.40&65.96&71.99&61.91&61.61&60.92&61.41&59.32&61.45\\
  paper4&47.64&58.33&59.69&60.95&68.46&56.35&56.12&55.30&55.75&53.62&55.80\\
  paper5&44.96&58.21&59.47&59.54&67.39&52.97&52.83&52.30&52.61&51.02&52.47\\
  paper6&50.94&65.27&67.30&67.74&75.06&61.72&61.50&60.84&61.26&59.56&61.18\\
  pic&87.88&89.00&92.04&90.30&95.61&90.06&90.11&90.33&90.29&90.22&90.31\\
  progc&51.67&66.49&68.42&68.33&75.99&61.51&61.39&60.79&61.13&59.70&61.08\\
  progl&62.11&77.29&79.18&78.26&85.18&71.58&71.47&71.15&71.46&70.10&71.53\\
  progp&61.10&77.23&79.13&78.31&85.56&70.73&70.57&69.96&70.35&69.03&70.29\\
  trans&59.19&79.74&82.25&80.90&87.58&70.76&70.61&69.94&70.35&68.89&70.32\\
  merge&57.95&67.15&73.82&72.63&80.94&66.11&67.15&67.40&67.00&67.36&67.28\\
  \end{tabular}
  }
\end{table}

\begin{table}[ht!]
  \caption{Compression result, Canterbury corpus}
  \label{table:canterbury-results}
  \resizebox{\textwidth}{!} {
  \begin{tabular}{r|c|c|c|c|c|c|c|c|c|c|c}
  file&LZW&gzip&LZMA&bzip2&paq8l&CTW&ACTW1&ACTW2&ACTW3&ACTW4&ACTW5\\
  \hline\hline
  alice29.txt&59.07&64.21&68.13&71.59&76.96&69.23&68.85&68.42&68.91&67.15&68.93\\
  asyoulik.txt&56.07&60.90&64.45&68.39&73.60&66.61&66.22&65.68&66.20&64.32&66.22\\
  cp.html&54.00&67.49&69.14&69.01&76.07&62.72&62.43&61.48&62.06&60.03&61.95\\
  fields.c&55.48&71.81&73.30&72.74&79.97&64.25&64.01&63.15&63.60&61.97&63.49\\
  grammar.lsp&51.28&66.51&66.60&65.52&74.12&58.18&58.08&57.43&57.70&56.49&57.46\\
  kennedy.xls&69.85&79.92&94.66&87.35&98.86&75.15&82.59&83.36&80.88&83.43&83.31\\
  lcet10.txt&61.77&66.05&72.02&74.76&80.34&70.18&69.83&69.62&70.00&68.67&70.00\\
  plrabn12.txt&59.12&59.49&65.68&69.79&74.69&68.37&67.88&67.74&68.19&66.65&68.13\\
  ptt5&87.88&89.00&92.04&90.30&95.61&90.06&90.11&90.33&90.29&90.22&90.31\\
  sum&47.43&66.20&75.36&66.24&80.33&57.14&57.51&57.25&57.10&57.03&57.10\\
  xargs.1&44.67&58.46&58.34&58.32&66.55&50.06&49.82&48.78&49.21&47.36&49.25\\
  \end{tabular}
  }
\end{table}

\begin{table}[ht!]
  \caption{Compression result, SFC testset}
  \label{table:sfc-results}
  \resizebox{\textwidth}{!} {
  \begin{tabular}{r|c|c|c|c|c|c|c|c|c|c|c}
  file&LZW&gzip&LZMA&bzip2&paq8l&CTW&ACTW1&ACTW2&ACTW3&ACTW4&ACTW5\\
  \hline\hline
  A10.jpg&0.00&0.12&-0.41&0.71&17.09&-3.86&-3.77&-3.47&-3.55&-3.24&-3.69\\
  AcroRd32.exe&37.79&55.19&63.65&56.09&76.05&46.02&47.83&48.19&47.53&48.71&47.82\\
  english.dic&62.69&74.18&79.06&69.96&90.47&65.18&75.87&76.67&74.43&76.14&76.74\\
  FlashMX.pdf&0.00&15.26&18.11&15.82&21.27&8.62&8.92&9.30&9.17&9.58&9.07\\
  FP.LOG&86.90&92.97&96.03&96.49&98.69&91.51&92.00&92.48&92.18&92.35&92.58\\
  MSO97.DLL&22.74&42.07&51.95&44.19&65.28&36.65&37.70&37.85&37.40&38.17&37.58\\
  ohs.doc&62.72&75.65&81.07&78.24&86.87&73.83&73.47&74.18&74.19&74.28&74.09\\
  rafale.bmp&65.18&69.60&76.50&78.55&84.71&75.69&76.84&77.15&77.04&76.53&77.40\\
  vcfiu.hlp&64.65&79.41&85.20&82.71&90.21&67.13&71.08&71.16&69.68&71.31&71.03\\
  world95.txt&62.07&70.78&80.95&80.69&87.81&70.37&70.37&70.42&70.56&69.88&70.64\\
  merge&55.01&71.56&76.68&74.53&82.77&64.30&67.86&68.39&67.30&68.73&68.14\\
  \end{tabular}
  }
\end{table}

\begin{table}[ht!]
  \caption{Compression result, Gauntlet benchmark}
  \label{table:gauntlet-results}
  \resizebox{\textwidth}{!} {
  \begin{tabular}{r|c|c|c|c|c|c|c|c|c|c|c}
  file&LZW&gzip&LZMA&bzip2&paq8l&CTW&ACTW1&ACTW2&ACTW3&ACTW4&ACTW5\\
  \hline\hline
  abac&99.45&99.88&99.94&99.98&99.95&99.98&98.73&99.59&99.91&99.57&99.63\\
  abba&95.12&94.16&99.79&96.16&98.21&96.43&95.21&96.32&96.42&96.28&96.36\\
  book1x20&60.27&59.39&98.29&71.83&98.71&69.97&69.21&69.44&69.80&69.13&69.63\\
  fib\_s14930352&99.22&99.27&99.85&99.99&99.96&95.42&94.12&95.31&95.41&95.28&95.35\\
  fss10&99.04&98.95&99.94&99.99&99.96&93.45&92.15&93.34&93.44&93.29&93.39\\
  fss9&98.51&98.93&99.94&99.98&99.96&93.45&92.15&93.27&93.43&93.18&93.36\\
  houston&98.96&98.75&99.22&99.39&99.54&88.95&88.13&88.82&88.93&88.80&88.84\\
  paper5x80&77.28&98.75&99.48&98.02&99.57&82.79&81.85&81.45&82.27&80.78&81.92\\
  test1&95.98&99.54&99.96&99.86&99.96&91.94&90.96&91.28&91.76&91.11&91.37\\
  test2&95.98&99.54&99.96&99.86&99.96&91.96&91.01&91.35&91.80&91.18&91.45\\
  test3&0.00&3.93&98.06&92.37&99.96&75.98&75.27&73.56&74.70&72.61&74.47\\
  merge&82.62&86.22&99.41&92.60&99.37&86.77&86.74&87.49&87.57&87.38&87.57\\
  \end{tabular}
  }
\end{table}

\begin{table}[ht!]
  \caption{Compression result, assorted files}
  \label{table:assorted-results}
  \resizebox{\textwidth}{!} {
  \begin{tabular}{r|c|c|c|c|c|c|c|c|c|c|c}
  file&LZW&gzip&LZMA&bzip2&paq8l&CTW&ACTW1&ACTW2&ACTW3&ACTW4&ACTW5\\
  \hline\hline
  book1.pdf&0.00&22.82&25.06&23.11&27.37&15.73&16.71&17.16&16.84&17.47&16.89\\
  book2.txt&60.75&62.82&71.39&73.41&79.61&69.99&69.65&69.80&70.04&69.22&70.05\\
  data1.xls&65.04&71.02&81.73&76.45&89.67&68.20&70.98&71.18&70.04&71.51&70.88\\
  data2.xls&64.85&73.26&85.17&76.96&93.53&66.95&69.64&69.90&68.64&70.24&69.66\\
  exec1&28.78&47.45&55.60&48.06&67.38&38.73&40.05&40.40&39.86&40.80&40.00\\
  exec2.exe&0.00&1.01&-0.07&0.22&1.52&-9.19&-9.05&-8.55&-8.70&-8.15&-8.87\\
  exec3.msi&0.00&4.60&4.27&3.91&6.00&-5.53&-5.28&-4.79&-4.96&-4.40&-5.06\\
  flash.swf&0.00&0.29&-0.73&-0.38&0.64&-9.86&-9.76&-9.35&-9.45&-9.04&-9.59\\
  foreign.hwp&0&5.78&6.08&4.68&7.31&-3.65&-3.57&-3.13&-3.22&-2.80&-3.41\\
  lib.dll&0.00&15.48&16.83&13.49&19.58&5.50&5.89&6.31&6.11&6.69&5.95\\
  pic1.png&0.00&0.19&-0.12&-0.30&3.37&-8.86&-8.75&-8.28&-8.40&-7.89&-8.61\\
  pic2.png&0.00&0.05&0.43&0.43&3.24&-6.99&-6.88&-6.49&-6.59&-6.16&-6.76\\
  pitches&37.91&69.71&74.00&64.27&78.15&49.60&52.72&52.49&50.93&52.77&52.45\\
  pres.ppt&0.00&61.72&66.14&62.12&74.81&49.86&50.13&50.00&50.01&49.94&49.78\\
  text.rtf&67.39&84.55&85.81&85.69&89.70&77.48&77.24&76.70&77.24&75.64&76.99\\
  vid1.avi&0.00&3.48&3.93&3.54&6.12&-1.02&-0.86&-0.49&-0.59&-0.22&-0.74\\
  vid2.mov&0.00&1.76&3.11&2.12&7.91&0.42&0.60&0.87&0.77&1.32&0.62\\
  merge1&0.00&20.74&23.47&20.92&27.59&11.72&12.83&13.59&13.04&14.33&13.02\\
  merge2&5.00&30.08&33.94&33.42&38.92&26.01&26.40&26.80&26.63&27.10&26.47\\
  merge3&0&7.52&7.70&6.85&9.95&-2.73&-2.39&-1.89&-2.11&-1.52&-2.17\\
  merge4&0.00&10.20&12.46&10.27&17.34&6.21&7.01&7.43&6.98&8.13&6.96\\
  \end{tabular}
  }
\end{table}

\paradot{Experimental results\footnote{The numbers in the tables are defined as $1- \frac{Compressed\ size}{Uncompressed\ size}$. The larger the better.}}
From analysis of these results the partial context visit based
adaptive CTW with parameters $ c = 0.1 $, $ \alpha = 0.33 $
(ACTW2) appeared to produce the best compression results. Note
that we compare generic bitwise compressors ACTW and CTW with
highly tuned ones. For example, bzip2 uses several layers of
compression techniques stacked on top of each other during
compression. This comparison is not entirely fair as CTW and
ACTW are not tuned. Therefore the results should not be taken
as an indication that ACTW and CTW are necessarily inferior
compression techniques.

Table \ref{table:calgary-results} shows the results for the
selected compressors on the large Calgary orpus. We see that
the adaptive modification leads to better compression in only
four test files, while for the remaining fourteen, the standard
CTW algorithm is slightly better. Nevertheless, the difference
exceeds 1\% (1.07\%) only in the bib case. Despite the general
trend  favouring the standard CTW approach, when compressing
the concatenation of the corpus, adaptive CTW gives an
improvement of nearly 1.3\%.

Results for the Canterbury Corpus in Table
\ref{table:canterbury-results} once again seems to generally
favour the standard CTW algorithm, though again only by small
amounts. One exception to this trend is the kennedy.xls file,
where adaptive CTW modifications allow the space savings to be
improved by over 8\%. For two compressors based on the same
principles this is a significant improvement.

The SFC test set results given in Table \ref{table:sfc-results}
show ACTW2 consistently outperforming standard CTW. Some of
these are quite significant increases, including an improvement
of over 2\% for AcroRd32.exe, 4\% for vcfiu.hlp and 10.5\% for
english.dic.

Once again, considering the close relation to the standard CTW
algorithm this is a very significant improvement in space
savings. English.dic is an alphabetically sorted word file. It
appears the adaptive modifications allow ACTW2 to better adapt
to the changing distribution of the text in the file. This
leads the ACTW2 to outperform bzip2 by around $5\%$ while the
standard CTW algorithm trailed bzip2 as much.

Again we see ACTW2 perform better when compressing the
concatenated test set. However as ACTW2 was generally
performing better on the individual files this is not as
indicative of how adaptation improves compression when using
varied sources. One other observation of interest here is how
ACTW2 is able to perform very closely to bzip2 for the
rafale.bmp file. Something similar was seen in the Calgary
corpus case, where ACTW2 gave better results than bzip2 for the
pic file, also a bitmap image. Consulting the index it can be
seen that this is not a special case of poor performance for
the bzip2 compressor, but instead seems to indicate some
property of the bitmap file format that makes it particularly
suited to CTW based compression.

In the Gauntlet benchmark results seen in Table
\ref{table:gauntlet-results} we once more see CTW outperforming
ACTW. While the margin between the two is quite small, CTW
gives better compression for every individual file in this test
set. However looking at the concatenated file results we see
ACTW2 giving around a 0.8\% improvement in space savings. As
this cannot be the result of better compression for any
individual file, it must indicate that the adaptive
modifications allow ACTW2 to better respond to changing input
distributions.

For the assorted file collection results given in Table
\ref{table:assorted-results} we see ACTW2 generally offering
better compression than standard CTW, including for all four
concatenated files tested here. Space saving increases of
around 3\% can be seen for both data1.xls and data2.xls. In
combination with the improvement of more than 8\% for
kennedy.xls in the Canterbury corpus this suggests there might
be something about the Microsoft Excel file format that makes
it well suited to ACTW2.

\section{Conclusion} \label{sec:conclusion}

We proposed the ACTW algorithm as an extension to the standard
CTW algorithm that puts greater weighting on more recent
observations. Four different versions of ACTW algorithms were
tested to determine the effects on prediction performance. The
performance of ACTW, especially partial context visit based
adaptive CTW with $ c = 0.1 $, $ \alpha = 0.33 $, was promising
with space saving improvements up to 10\% when compared to the
standard CTW algorithm. While the standard CTW was able to
outperform ACTW for a number of files, the difference in space
savings for these cases very rarely exceeded 1\%. For
concatenated files from varying sources, ACTW was seen to
outperform standard CTW for all tests performed, demonstrating
how the adaptive modifications to CTW allow better handling of
changing source distributions. This improved performance for
concatenated files was even observed when the standard CTW
algorithm provided better compression for each of the
individual files.

\addcontentsline{toc}{section}{\refname}

\begin{footnotesize}
  \providecommand{\bysame}{\leavevmode\hbox to3em{\hrulefill}\thinspace}
  \providecommand{\MR}{\relax\ifhmode\unskip\space\fi MR }
  \providecommand{\MRhref}[2]{%
    \href{http://www.ams.org/mathscinet-getitem?mr=#1}{#2}
  }
  \providecommand{\href}[2]{#2}
  
\end{footnotesize}

\end{document}